\documentclass[aps,showpacs,amsmath]{revtex4}

\usepackage{graphicx}

\begin{document}
\title{Ion shock acceleration by large amplitude slow ion acoustic double layers in laser-produced plasmas}
\author{B. Eliasson}
\affiliation{SUPA, Physics Department, John Anderson Building, University of Strathclyde,
Glasgow G4 0NG, UK}
\begin{abstract}
A kinetic model for the shock acceleration of ions in laser-produced plasmas
is developed. A fraction of the warm ions are accelerated by the
large amplitude monotonic potential of the shock created due
the plasma compression and electron heating by the laser.
The kinetic model for the monotonic shock is based
on the slow ion acoustic double layer (SIADL).
It is found that the amplitude of the large amplitude SIADL is almost uniquely
defined by the electron temperature. Therefore, a balance between electron heating
and plasma compression is needed for optimal ion acceleration by this scheme.
Typical Mach numbers of the monotonic shocks are close to 1.5.
The scheme could potentially produce monoenergetic ions with a relative
energy spread of less than 1 percent.
The model is compared with recent simulations and experiments, where efficient
shocks acceleration and production of monoenergetic protons have been observed.
Similarities and differences with other shock models are pointed out and discussed.
\end{abstract}
\pacs{52.38.Kd,52.35.Tc,52.25.Dg}
\received{8 January 2013}
\revised{5 February 2014}
\maketitle

\section{Introduction}
Laser accelerated ions have important potential applications in the fields of
cancer therapy \cite{Bulanov02b, Ledingham07, Linz07},
inertial confinement fusion \cite{Roth01}, etc., where well defined
energies of the protons are of great importance. A number of ion acceleration
methods have been considered \cite{Mora07,Macci13,Sgattoni13}.
The interaction of an intense laser beam with an underdense plasma can lead to rapid heating
of the electrons and acceleration of ions \cite{Krushelnick99,Wei04}.
In target normal sheath acceleration (TNSA) \cite{Sentoku00,Hatchett00,Wilks01,Bulanov02,Dong03,Zhou07,Robson07,Lee08,Passoni08,Antici09},
a minority species of light ions (typically protons)
is accelerated by the hot sheath of laser-heated electrons normal to a
solid target. The TNSA scheme has successfully been used to
accelerate protons to several tens of MeV, usually with a wide spectrum of
energies \cite{Passoni08}.
In radiation pressure acceleration (RPA)
\cite{Esirkepov04, Yan08, Klimo08,Robinson08,Tripathi09,Eliasson09},
a thin target is accelerated by the radiation pressure of a circularly polarized
laser, leading to a more peaked energy spectrum of the ions. The most severe
problem that so far has prevented a greater success of the RPA scheme is
the Rayleigh-Taylor instability \cite{Ott72,Pegoraro07}, which can rapidly disrupt the foil and
broaden the ion energy spectrum \cite{Liu11}. Multi-ion species foils combined
with different polarizations have been used to
improve the energy spectrum of the accelerated protons \cite{Liu13,Liu13b}.
Shock-related mechanisms
for reflection and acceleration of ions in the contexts of space
and laboratory plasmas have been investigated
in the past \cite{Sagdeev66,Forslund71,Tidman71,Perkins81,Michelsen82}, and
go under names such as laminar electrostatic shocks, double layers, etc.
Different types of double layers have been treated theoretically by Schamel and co-workers
\cite{Schamel82,Schamel83,Schamel86}, including slow ion acoustic double layers
(SIADLs) associated with the nonlinear slow ion acoustic mode,
slow electron acoustic double layers (SEADLs)associated with
the nonlinear slow electron acoustic mode, strong double layers \cite{SB83}
with beam-like free and trapped particle distributions, and current-free double layers \cite{Goswami08}.
A different scheme that has gained
interest is combined radiation pressure and shock or soliton acceleration of ions
\cite{Silva04,Palmer11,Haberberger12,He12,Cairns13, Macci12}. In this scheme, the laser is
injected towards an over-dense gas target, which leads to a
compression of the plasma and heating of the electrons. This can result in
a large amplitude shock-like or soliton-like structure \cite{Macci12}
that propagates into the target.
Due to the large amplitude electrostatic
potential, a portion of the upstream ions are specularly
reflected by the nonlinear structure. Since the
target is over-dense, the shock or soliton does not interact directly
with the laser light after the initial compression of the plasma.

Recent kinetic shock models either
treat the ions as a cold fluid \cite{Sorasio06,Fiuza12} or kinetically
with a finite momentum spread \cite{Stockem13,Fiuza13}, while the electrons are
divided into free and trapped/reflected downstream populations.
Shocks with oscillatory downstream shock potentials
have been found by including hot ions and considering
trapped downstream electron populations and a population
of reflected upstream ions \cite{Cairns13,Stockem13,Fiuza13}.
The aim is here to develop a kinetic model of monotonic (non-oscillatory) shock
structures in un-magnetized plasmas, including hot electrons and ions, and to
investigate the conditions for the existence of monotonic shocks
on the plasma parameters.

\section{Kinetic shock model}
Our model covers the structure
of the shock, and how the ions are reflected by the self-consistent
large amplitude potential.  A solution of the
time-independent ion Vlasov equation
\begin{equation}
  v\frac{\partial f_i}{\partial x}-\frac{\partial \phi}{\partial x}\frac{\partial f_i}{\partial v}=0
  \label{eq1}
\end{equation}
is given by $f_i({\cal E})$ where
\begin{equation}
  {\cal E}=\frac{v^2}{2}+\phi-\phi_{\rm max},
  \label{eq2}
\end{equation}
is the conserved energy of an ion moving with velocity $v$
under the influence of the electrostatic potential $\phi$.
We assume that the potential is at $\phi=\phi_{\rm max}$ at $x=-\infty$,
and drops at the double layer so that $\phi=0$ at $x=+\infty$, where the ion distribution
contains reflected ions.
In Eqs. (\ref{eq1}) and (\ref{eq2}), the velocity $v$ has been normalized by the ion thermal velocity $v_{Ti}=\sqrt{k_B T_i/m_i}$, the potentials $\phi$ and $\phi_{\rm max}$ by $k_B T_i/e$, space $x$ by $v_{Ti}/\omega_{pi}$ where $\omega_{pi}=\sqrt{n_0e^2/(\epsilon_0 m_i)}$
is the ion plasma frequency, and the ion distribution function $f_i$ by
$n_0/v_{Ti}$. Here $T_i$ and $n_0$ is the ion temperature and number density, respectively, at
$x=-\infty$. Furthermore, $k_B$ is Boltzmann's constant, $\epsilon_0$ is the electric vacuum permittivity,
$e$ is the magnitude of the elementary charge, and $m_i$ is the ion mass.
Ions with ${\cal E}>0$ are ``free'' and can penetrate the shock, while ions
with ${\cal E}<0$ are reflected by the shock potential.
Our model is based on the one-dimensional Vlasov-Poisson system using
Schamel's theory of SIADLs \cite{Schamel86} to construct well-behaved distribution functions.
In the frame of the shock, moving with a velocity $u_{i0}$ along the $x$-axis,
the equilibrium ions at $x=-\infty$ obey a shifted Maxwellian distribution function $f_i=(1/\sqrt{2\pi})\exp[-(v+u_{i0})^2/2]$
streaming in the opposite direction. The particular choice of downstream ion distribution
seems to be quite consistent with the recent simulation results of laser produced shocks in
Ref. \cite{Haberberger12} (see their Fig. 4c-d).
This is fulfilled by the special choice of
ion velocity distribution function \cite{Schamel72,Schamel86}
\begin{equation}
  f_i=\frac{1}{\sqrt{2\pi}}\left\{ \begin{array}{ll}
  \exp\left[-\bigg(\sqrt{{\cal E}}+\frac{u_{i0}}{\sqrt{2}}\bigg)^2\right], & v>v_{+},
  \\
  \exp\left[-\bigg( \alpha{\cal E}+\frac{u_{i0}^2}{2}\bigg)\right],
   & v_{-}\leq v \leq v_{+},
  \\
  \exp\left[-\bigg(-\sqrt{{\cal E}}+\frac{ u_{i0}}{\sqrt{2}}\bigg)^2\right], & v<v_{-},
  \end{array}
  \right.
  \label{eq_fi}
\end{equation}
where $v_\pm=\pm 2^{1/2}(\phi_{\rm max}-\phi)^{1/2}$ defines the limit
${\cal E}=0$ between the reflected (${\cal E}<0$) and free (${\cal E}>0$) ions in velocity space,
and $u_{i0}$ is the velocity of the double layer compared to the mean velocity of the ions at $x=-\infty$.
The free ions are divided into two populations, one which overtakes the shock ($v>v_+$), and
one which is overtaken by the shock ($v<v_-$), hence $f_i$ is double-valued for ${\cal E}>0$.
The reflected ions in the range $v_-\geq v\geq v_+$ are characterized
by the parameter $\alpha$, where $\alpha=0$ gives a flat-topped reflected
ion population, $\alpha<0$ gives a beam-like, excavated population in velocity space, and
$\alpha=-\infty$ gives a completely depleted distribution with no reflected ions.
The parameter $\alpha$ is sometimes referred to as the ``trapping parameter'' in the context of ion holes \cite{Schamel86} to describe trapped particle distributions.
The ion density $n_i$ (normalized by $n_0$) is obtained by integrating
Eq.~(\ref{eq_fi}) over velocity space and can be expressed in terms of special functions as \cite{Schamel82,Bujarbarua81}
\begin{equation}
  n_i(\phi,\phi_{\rm max})=\exp\bigg(-\frac{u_{i0}^2}{2}\bigg)
  \bigg\{ I(\phi_{\rm max}-\phi)+K\bigg(\frac{u_{i0}^2}{2},\phi_{\rm max}-\phi\bigg)
  +\frac{2}{\sqrt{\pi|\alpha|}}W\big[\sqrt{\alpha(\phi-\phi_{\rm max})}\big]
  \bigg\}
\end{equation}
where $I(\Phi)=\exp(\Phi)[1-{\rm erf}(\sqrt{\Phi})]$, ${\rm erf}(u)$ is the error function,
\begin{equation}
  K(X,\Phi)=\frac{2}{\sqrt{\pi}}\int_0^{\pi/2} \sqrt{X}\cos(\Phi)\exp(-\Phi\tan^2\theta+X\cos^2\theta){\rm erf}(\sqrt{X}\cos\theta)\,d\theta \,,
\end{equation}
and
\begin{equation}
  W(u)=\exp(-u^2)\int_0^u \exp(t^2)\,dt
\end{equation}
is Dawson's integral.
The electrons are assumed to be isothermal
so that the electron number density obeys the Boltzmann
distribution of the form \cite{Cairns13}
\begin{equation}
  n_e(\phi,\phi_{\rm max})=n_i(0,\phi_{\rm max})\exp(\phi/\theta),
  \label{eq_ne}
\end{equation}
where $\theta=T_e/T_i$. Equation (\ref{eq_ne}) ensures that $n_e=n_i$ when $\phi=0$.
It should be noted that kinetic models for the electrons have been used both in the
non-relativistic \cite{Forslund71,Sorasio06} and
relativistic \cite{Stockem13,Fiuza13} regimes; to obtain Eq.~(\ref{eq_ne}),
we have made the simplifying assumption that the electrons are thermalized on the ion time-scale and
assume a non-relativistic Maxwell-Boltzmann distribution.
Inserting the expressions for the ion and electron densities into Poisson's equation,
we have
\begin{equation}
  \frac{d^2\phi}{dx^2}=n_e(\phi,\phi_{\rm max})-n_i(\phi,\phi_{\rm max}).
\end{equation}
By defining the Sagdeev potential
\begin{equation}
   V(\phi,\phi_{\rm max})=\int_0^\phi[n_i(\phi',\phi_{\rm max})-n_e(\phi',\phi_{\rm max})]\,d\phi'
\label{eq_psi}
\end{equation}
we have from Poisson's equation
\begin{equation}
  \frac{d^2\phi}{dx^2}=-\frac{\partial V}{\partial \phi},
\end{equation}
which, multiplying by $d\phi/dx$, gives
\begin{equation}
  \frac{1}{2}\bigg(\frac{d\phi}{dx}\bigg)^2=-V.
  \label{sagdeev}
\end{equation}
For the existence of a laminar shock, the conditions are that both $V=0$ and
$\partial V/\partial \phi=0$ for $\phi=0$ and $\phi=\phi_{\rm max}$, and
that $V<0$ for $0<\phi<\phi_{\rm max}$. In particular, the condition
\begin{equation}
 V(\phi_{\rm max},\phi_{\rm max})=0
  \label{cond1}
\end{equation}
defines the amplitude of the shock, while it holds trivially from Eq.~(\ref{eq_psi})
that $V(0,\phi_{\rm max})=0$.
The condition $\partial V/\partial \phi=0$ at
$\phi=\phi_{\rm max}$ is equivalent to the quasi-neutrality condition $n_e=n_i$ at $x=-\infty$,
where it also holds that $n_i(\phi_{\rm max},\phi_{\rm max})=1$.
Using this together with the expression (\ref{eq_ne}) for the electron density, we have
\begin{equation}
  n_i(0,\phi_{\rm max})=\exp(-\phi_{\rm max}/\theta).
  \label{cond2}
\end{equation}
The conditions (\ref{cond1}) and (\ref{cond2}) give $\phi_{\rm max}$ and
$u_{i0}$ for given values of $\theta$ and $\alpha$. After these have been found, Eq.~(\ref{sagdeev})
can be integrated to obtain the spatial profile of the shock potential.
It should be noted that the charge neutrality condition (\ref{cond2}) leads to monotonic shocks, in contrast
 to the models of oscillatory shocks \cite{Cairns13,Stockem13,Fiuza13} where
 different boundary conditions were used.

\begin{figure}[htb]
\centering
\includegraphics[width=18cm]{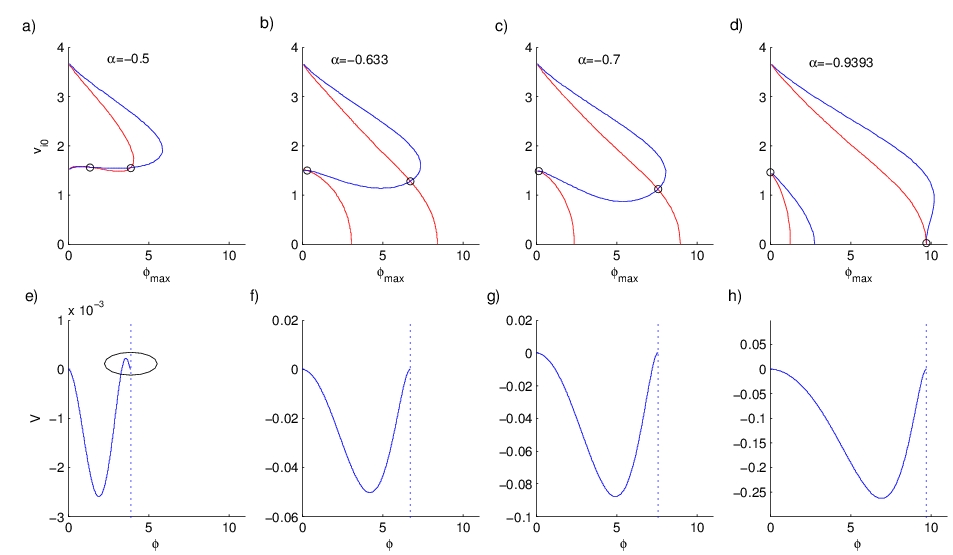}
\caption{(color online) a)--d): Solutions of Eqs.~(\ref{cond1}) and (\ref{cond2}) (blue and red curves, respectively) in the
($\phi_{\rm max}$, $u_{i0}$)-plane for $\theta=10$ and different values of $\alpha$.
Where the curve intersect each other (indicated with circles),
both equations are simultaneously solved, indicating possible double layer solutions.
The large $\phi_{max}$ solutions are
a) $(u_{i0},\phi_{\rm max})= (1.5467,\,3.8935)$ for $\alpha=-0.5$,
b) $(u_{i0},\phi_{\rm max})= (1.2785,\,6.7222)$ for $\alpha=-0.633$,
c) $(u_{i0},\phi_{\rm max})=(1.1149,\,7.5699)$ for $\alpha=-0.7$, and
d) $(u_{i0},\phi_{\rm max})=(0.0147,\,9.70235)$ for $\alpha=-0.9393$.
The corresponding Sagdeev potentials are shown in panels e)--f).
The vertical dotted lines indicate $\phi=\phi_{\rm max}$.
For $\alpha>-0.633$ the shock solution is not permitted, since in this case $V>0$ part of
the interval $0<\phi<\phi_{\rm max}$ [e.g.~in panel e) $\alpha=-0.5$ indicated by an ellipse], while $-0.633>\alpha>-0.9393$
permit shock solutions [panels f)--h)].
}
\end{figure}



\begin{figure}[htb]
\centering
\includegraphics[width=18cm]{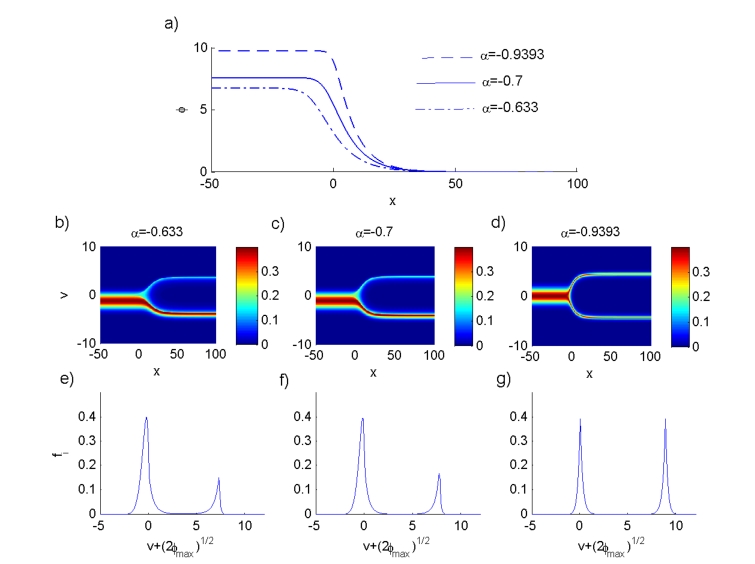}
\caption{a) Profiles of the electrostatic potential for $\theta=10$ and $\alpha=-0.633$ (dash-dotted curve),
$\alpha=-0.7$ (solid curve), and $\alpha=-0.9393$ (dashed curve). b)--d): The ion distribution function $f_i(x,v)$ for b) $\alpha=-0.633$, c) $\alpha=-0.7$, and d) $\alpha=-0.9393$.
e)--g): The corresponding upstream ion distribution function in the frame of the incoming plasma.
A beam is seen with a relative speed of approximately $2(2\phi_{\rm max})^{1/2}$, corresponding
to an ion energy of about $4\phi_{\rm max}$ relative to the upstream ions, which can be assume to be
at rest in the laboratory frame.  The relative energy spread of the beam is $1.0\times10^{-2}$, $6.3\times10^{-3}$
and $1.7\times10^{-3}$ for the respective cases e)-g), with a relative beam density of 0.24, 0.29 and 1.0, respectively, compared
to the upstream ion density.}
\end{figure}


\section{Numerical results}
Figures 1a--d show solutions of Eqs.~(\ref{cond1}) and (\ref{cond2}) in the ($u_{i0}$, $\phi_{\rm max}$)-plane
for $\theta=10$ and different values of $\alpha$. Simultaneous solutions of Eqs.~(\ref{cond1}) and (\ref{cond2})
are found where the two curves intersect each other.
These solutions are associated with the nonlinear slow ion acoustic mode (discussed below), and
the corresponding double layers are referred to as slow ion acoustic double layers
(SIADLs) in the classification scheme of Schamel \cite{Schamel82,Schamel83,Schamel86}.
We see in Figs.~1a--d that the system has solutions for both large and small values of $\phi_{\rm max}$,
which could indicate the existence of double layers.  However, the small amplitude
solutions seen in Figs.~1a--d (for $\phi_{\rm max}<2$) produce $V>0$ in the interval $0<\phi<\phi_{\rm max}$
and hence do not permit double layers. The existence of small amplitude, weak double layers
requires non-Maxwellian electrons which are more flat-topped or excavated \cite{Kim83,Schamel83}.
The value $\alpha=-0.9393$ is close to
the maximum negative limit, beyond which no large amplitude solutions exist.
The double layers can be seen as limiting cases of ion hole solutions,
(see Fig. 6 of Ref. \cite{Bujarbarua81}), where
the maximum negative $\alpha$ corresponds to a current-free double layer.
The condition
that $V<0$ in the interval $0<\phi<\phi_{\rm max}$ further restricts
the existence of double layers. It turns out that $\alpha$ must have larger negative
values than about $-0.633$ for laminar shocks to exist, as illustrated in Figs.~1e--h. For
smaller values, such as $\alpha=-0.5$ (cf.~Fig.~1e), $V$ is larger than zero in part
of the interval and prevents the existence of double layers. Equations (\ref{cond1}) and (\ref{cond2})
also have simultaneous solutions in the linear limit $\phi_{\rm max}\rightarrow 0$.
In this limit, the phase velocity $u_{i0}=v_{ph}$ is given by the dispersion relation for undamped ion acoustic waves
in the long-wavelength limit (omitting Landau damping by keeping only the principal part of the
Landau contour integral),
\begin{equation}
  1-\pi v_{ph} H[f_{i0}(v_{ph})]+1/\theta=0
  \label{disp_rel}
\end{equation}
where
\begin{equation}
  H[f_{i0}(v_{ph})]=-\frac{1}{\pi}\int_0^\infty\frac{f_{i0}(v_{ph}+u)-f_{i0}(v_{ph}-u)}{u}\,du
\end{equation}
is the Hilbert transform of the Maxwellian equilibrium ion distribution function $f_{i0}(v)=(2\pi)^{-1/2}\exp(-v^2/2)$
\cite{Fried61}.
The dispersion relation has one high-velocity solution associated with the
usual linear ion acoustic mode $v_{ph}\approx \sqrt{3+\theta}$, and one
low-velocity solution associated with the slow ion acoustic mode \cite{Schamel80,Schamel86}
$v_{ph}\approx 1.305(1+\theta^{-1})$, for $\theta\gg 1$.
For $\theta=10$ we find numerically from the dispersion relation (\ref{disp_rel}) and
from Figs.~1a--d that
the velocity of the slow mode is given by $v_{ph}=1.45$ and the fast ion acoustic mode $v_{ph}=3.72$.
The fast and slow ion acoustic modes merge for $\theta\approx 3.5$, and for smaller values of $\theta$ the undamped ion
acoustic modes cease to exist (see e.g.~Ref.~\cite{Schamel80}). It should be noted that
the existence criterion for finite amplitude ion holes and SIADLs are \cite{Bujarbarua81}
$u_{i0}<1.305(1+\theta^{-1})$, $\theta>3.5$, and $\alpha<0$ (strictly negative).



The spatial profiles of the electrostatic shock potential are plotted in Fig.~2a for the cases $\alpha=-0.633$, $-0.7$ and $-0.9393$.
The laminar shocks exhibit a jump from its high voltage value close to $\phi_{\rm max}$
at the left boundary to near zero at the right boundary. The corresponding
ion distribution functions, shown in Figs.~2b--d, are close to Maxwellian
at the left boundary, while it splits into two populations of ions at the right boundary
with upstream, incoming ions with negative velocities and a significant portion
of reflected ions with positive velocities.
The limiting large amplitude value $\alpha=-0.9393$ gives rise to two symmetric ion populations with equal incoming
and reflected densities.
In Figs.~2e--g, we show the ion velocity distribution at the right-hand boundary,
shifted to a frame moving with the upstream incoming ions (with negative velocities in Figs.~2b--d).
The reflected beam has a velocity close to $2(2\phi_{\rm max})^{1/2}$,
corresponding to an ion kinetic energy of $4\phi_{\rm max}$.
It should be emphasized that only large amplitude shock solutions exist, for a finite range of $\alpha$ in the
interval $-0.633>\alpha>-0.9393$ and with a corresponding amplitude of the shock potential in the range
$6.72<\phi_{\rm max}<9.7$. The shock speed (the downstream bulk speed minus the upstream, incoming beam speed) can be estimated as $v_{shock}=\sqrt{2\phi_{\rm max}}-u_{i0}$
and lies in the range $2.38<v_{shock}<4.4$. Dividing with the linear ion acoustic speed $v_{ph}\approx\sqrt{\theta}\approx 3.2$
gives a Mach number in the range $0.75<M<1.4$. The relative energy spread of the beam ions is about 1\% or less.


\begin{figure}[htb]
\centering
\includegraphics[width=18cm]{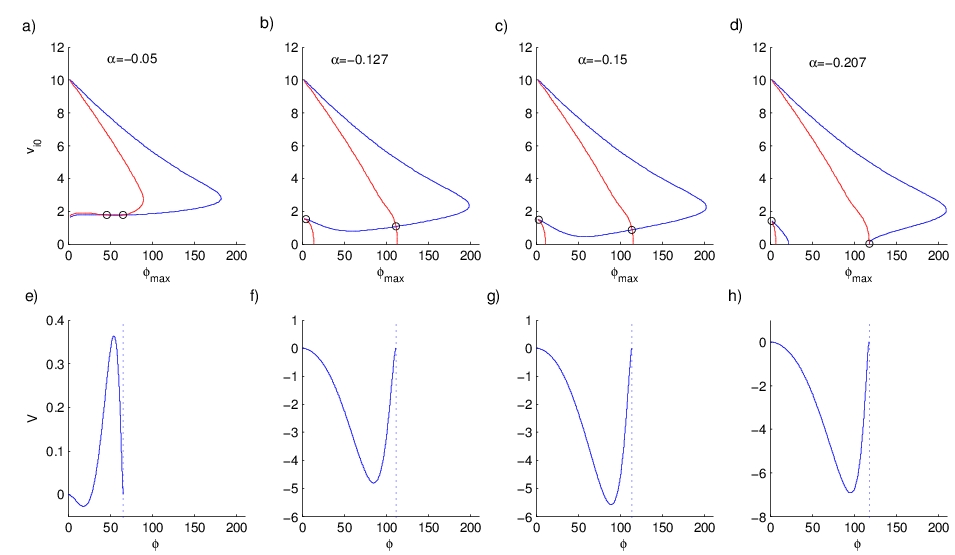}
\caption{a)--d): Solutions of Eqs.~(\ref{cond1}) and (\ref{cond2}) (blue and red curves, respectively) in the
($\phi_{\rm max}$, $u_{i0}$)-plane for $\theta=100$ and different values of $\alpha$.
Where the curve intersect each other (indicated with circles),
both equations are simultaneously solved, indicating possible double layer solutions.
The large $\phi_{max}$ solutions are
a) $(u_{i0},\phi_{\rm max})= (1.7651,\,65.3327)$ for $\alpha=-0.05$,
b) $(u_{i0},\phi_{\rm max})= (1.0686,\,111.5314)$ for $\alpha=-0.127$,
c) $(u_{i0},\phi_{\rm max})=(0.87084,\,113.8435)$ for $\alpha=-0.15$, and
d) $(u_{i0},\phi_{\rm max})=(0.022987,\,117.5444)$ for $\alpha=-0.207$.
The corresponding Sagdeev potentials are shown in panels e)--f).
The vertical dotted lines indicate $\phi=\phi_{\rm max}$.
For $\alpha>-0.127$ the shock solution is not permitted, since in this case $V>0$ part of
the interval $0<\phi<\phi_{\rm max}$ [e.g.~in panel e) $\alpha=-0.05$], while $-0.127>\alpha>-0.207$
[panels f)--h)] allow shock solutions.
}
\end{figure}

\begin{figure}[htb]
\centering
\includegraphics[width=18cm]{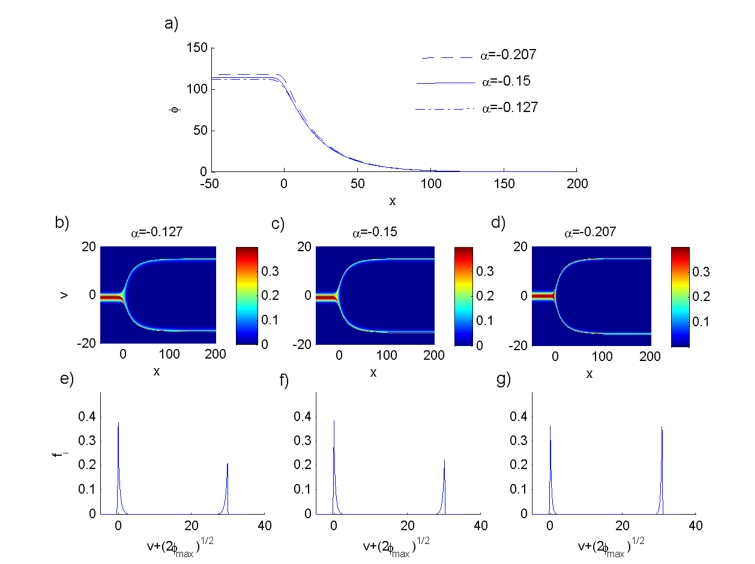}
\caption{a) Profiles of the electrostatic potential for $\theta=100$ and $\alpha=-0.127$ (dash-dotted curve),
$\alpha=-0.15$ (solid curve), and $\alpha=-0.207$ (dashed curve). b)--d):
The ion distribution function $f_i(x,v)$ for b) $\alpha=-0.127$, c) $\alpha=-0.15$, and d) $\alpha=-0.207$.
e)--g): The corresponding upstream ion distribution function in the frame of the incoming plasma.
A beam is seen with a relative speed of approximately $2(2\phi_{\rm max})^{1/2}$, corresponding
to an ion energy of about $4\phi_{\rm max}$ relative to the upstream ions, which can be assume to be
at rest in the laboratory frame. The relative energy spread of the beam is $4.1\times10^{-4}$, $2.6\times10^{-4}$
and $1.24\times10^{-4}$ for the respective cases e)-g), with a relative beam density of 0.65, 0.67 and 1.0, respectively, compared
to the upstream ion density.}
\end{figure}

Recent experiments and simulation studies \cite{Haberberger12,He12} have shown efficient shock acceleration of
protons using thick gas targets. The plasma is initially compressed by radiation pressure acceleration, the
electrons are heated, and a shock is formed propagating into the overdense plasma. A beam of reflected
ions with a narrow energy spectrum is subsequently formed. In the simulation study of Haberberger
et al.~\cite{Haberberger12}, the estimated electron temperature was about 100 times larger than the
upstream ion temperature, corresponding to $\theta=100$. In their Fig. 4, they observed
a significant fraction of reflected ions. The upstream, expanding ions had a velocity
of $0.1c$ relative to the laboratory frame, and the shock-reflected beam had a
velocity of $0.2c$, corresponding to an ion kinetic energy of about $20$ MeV relative
to the laboratory frame. As a comparison with these studies, show in
Figs. 3a--d solutions of Eqs.~(\ref{cond1}) and (\ref{cond2}) in the ($u_{i0}$, $\phi_{\rm max}$)-plane
for $\theta=100$ and different values of $\alpha$. We see in Figs.~3a--d that the system again
has solutions for both large and small values of $\phi_{\rm max}$,
of which it turns out that the small amplitude solutions do not support double layers
since $V$ takes positive values in the interval $0<\phi<\phi_{\rm max}$. The value $\alpha=-0.207$ is close to
the maximum amplitude limit, above which no shock solutions exist, and $\alpha$ must have
larger negative values than about $-0.127$ for laminar shocks to exist; see Figs.~3f--h. For
smaller values, such as $\alpha=-0.05$ (cf.~Fig.~4e), $V$ is larger than zero in part
of the interval and prevents the existence of double layers.



The associated profiles of the electrostatic potential are plotted in Fig.~4a,
and the ion distribution function is plotted in Figs.~4b--d where it is seen that a significant
portion ions are reflected by the shock. Here, the limiting large amplitude
value $\alpha=-0.207$ gives rise to two symmetric populations of incoming and reflected
ions with equal densities.
In the frame moving with the upstream ions, shown in Figs. 4e--g,
the reflected beam has a velocity close to $2(2\phi_{\rm max})^{1/2}$,
corresponding to an ion kinetic energy of $4\phi_{\rm max}$.
Only large amplitude solutions exist, for a finite range of $\alpha$ in the
interval $-0.127>\alpha>-0.207$ with a corresponding amplitude of the shock potential in the range
$111.5<\phi_{\rm max}<117.5$. Using an electron temperature of 1 MeV and an ion temperature of 10 keV,
 the ion beam energy $\approx 4\phi_{\rm max}\approx 450$ in dimensionless units corresponds
 in dimensional units to about 4.5 MeV and a relative beam speed of $0.07c$, calculated in the frame of the upstream ions.
In a laboratory frame where the upstream, expanding ions have velocity of $0.1c$,
the beam speed is $0.17c$ and the energy of the reflected ions are about 13 MeV.
This is comparable to, although somewhat lower than, the simulation results in Ref.~\cite{Haberberger12}
where ion energies up to 20 MeV were observed.
The higher ion energies in the simulations could be explained by an additional
sheath acceleration of the ions due to a finite size of the plasma, which is not taken
into account in the present theoretical model.
We observe that the shock speed $v_{shock}\approx \sqrt{2\phi_{max}}-u_{i0}$ is in the narrow range
$14.0<v_{shock}<15.3$, which gives a Mach number $1.4<M<1.5$ using
the linear ion acoustic speed $v_{ph}\approx\sqrt{\theta}=10$. Here the relative energy spread of the beam ions is
less than 0.1\%.

An interesting observation is that the monotonic shocks have both a \emph{maximum} and \emph{minimum} amplitude
for a given value of the electron-to-ion temperature ratio $\theta$. For large values of $\theta$,
the shock potential can only take a narrow range of large-amplitude values, as seen in Figs. 3 and 4.
To a good approximation, we have $\phi_{\rm max}/\theta\approx 1$ for $\theta$ ranging from $10$
to $100$. Using the shock velocity $v_{shock}\approx \sqrt{2\phi_{max}}\approx \sqrt{2\theta}$ and
dividing with the linear ion acoustic speed $v_{ph}\approx\sqrt{\theta}$ gives the Mach number $M\approx \sqrt{2}\approx 1.4$.
In dimensional units, the shock potential $\phi_{\rm max}$ is approximately equal to
$k_B T_e/e$ for the existence of the monotonic shock. The ratio of the upstream to the downstream electron density
is close to $\exp(e\phi_{\rm max}/k_B T_e)\approx 2.7$, and, since the upstream ions consist of
 and almost equal amount of reflected and incoming ions, the ratio between the downstream (incoming) ion density
to the upstream ion density is approximately 5.4. This determines the optimal rate of compression of
the plasma for the existence of the laminar shock. Hence, for laser compression, there should be a balance
$P_{rad}=P_e$, where $P_{rad}=2I_0/c$ is the radiation pressure ($I_0$ is the laser intensity and
$c$ the speed of light) and $P_e=n_0 k_B T_e$ is the electron pressure, taking into account that $n_0$
is about 5--6 times larger than the downstream density. If the laser amplitude is too small, there will not
be a well-defined double layer, since there is a predicted minimum amplitude of the double layer.
If the laser amplitude is too large, the laser will work as a piston and will always be in contact with the
shock front, similar to the cold ion piston case discussed in Ref. \cite{Forslund71}, and no double layer
will develop.

\section{Summary}

In summary, we have presented a kinetic model for the shock
acceleration of ions in laser-produced plasmas, where the
large amplitude laser is heating the electrons and compressing the overdense plasma
via the radiation pressure. The large amplitude shock specularly
reflects a large amount of the downs-stream ions.
It is found from the kinetic model that the amplitude of the
large amplitude shock is almost uniquely
defined by the electron temperature via $\phi_{\rm max}\approx k_B T_e/e$,
leading to an approximate ratio of 5--6 between the upstream and downstream ion densities.
Hence the model predicts an optimal balance between plasma compression and electron
heating for mono-energetic ion beams to be produced. The balance between
the laser radiation pressure and the downstream electron thermal pressure
then gives the optimal laser intensity for producing monoenergetic (potentially
less than 1\% energy spread) ions via
monotonic double layers/shocks. Typical Mach numbers of the monotonic shock are close to $M\approx 1.5$.
The analytic model agrees reasonably well with recent simulations and experiments \cite{Haberberger12}, where efficient
shocks acceleration and production of almost monoenergetic protons have
been observed using a linearly polarized laser focused into a supersonic hydrogen gas jet.
A recent experiment \cite{Palmer11} has also produced monoenergetic
protons of 4\% energy spread using circularly polarized laser and a hydrogen gas target.

\acknowledgments
Discussions with Bob Bingham (Rutherford Appleton Laboratory) and Francesco Pegoraro (University of Pisa) are gratefully acknowledged. The helpful comments by the anonymous Referee are appreciated.

\end{document}